\documentclass[conference]{IEEEtran}

\IEEEoverridecommandlockouts
\usepackage{verbatim}
\usepackage{longtable}
\DeclareMathAlphabet{\mathpzc}{OT1}{pzc}{m}{it}
\usepackage{amsmath}
\usepackage{amsfonts}
\usepackage{amssymb}
\usepackage{multirow}
\usepackage{amsmath,amssymb,amsfonts}

\usepackage[ruled,vlined,linesnumbered]{algorithm2e}
\usepackage[
top    = 0.7in,
bottom = 0.95in,
left   = 0.65in,
right  = 0.57in]{geometry}
\usepackage{graphicx}
\usepackage{textcomp}
\usepackage{xcolor}
\usepackage{subcaption}
\usepackage{algpseudocode}
\usepackage{enumitem}
\usepackage{float}
\restylefloat{figure}
\usepackage{graphicx}
\usepackage[normalem]{ulem}
\useunder{\uline}{\ul}{}
\usepackage{fancyhdr}

\begin{document}

\title{\textbf{ConChain}: A Scheme for \textbf{Con}tention-free and Attack Resilient Block\textbf{Chain}}

\author{\IEEEauthorblockN{Faisal Haque Bappy$^{1}$, Tariqul Islam$^{2}$,  Tarannum Shaila Zaman$^{3}$, \\ Md Sajidul Islam Sajid$^{4}$, and Mir Mehedi Ahsan Pritom$^{5}$}
\IEEEauthorblockA{
$^{1, 2}$ School of Information Studies (iSchool), Syracuse University, Syracuse, NY, USA\\
$ ^{3}$ Computer and Information Science, SUNY Polytechnic Institute, NY, USA\\
$ ^{4}$ Computer and Information Sciences, Towson University, Towson, MD, USA\\
$ ^{5}$ Computer Science, Tennessee Tech University, Cookeville, TN, USA\\
Email: \{fbappy@, mtislam@\}syr.edu and \{zamant@sunypoly, msajid@towson, mpritom@tntech\}.edu} 
}

\maketitle

\thispagestyle{fancy}
\lhead{This work has been accepted at the 2024 IEEE Consumer Communications \& Networking Conference (CCNC 2024)}
\cfoot{}

\pagestyle{empty}

\begin{abstract}

Although blockchains have become widely popular for their use in cryptocurrencies, they are now becoming pervasive as more traditional applications adopt blockchain to ensure data security. Despite being a secured network, blockchains have some tradeoffs such as high latency, low throughput, and transaction failures. One of the core problems behind these is improper management of “conflicting transactions”, which is also known as “contention”. When there is a large pool of pending transactions in a blockchain and some of them are conflicting, a situation of contention occurs, and as a result, the latency of the network increases, and a substantial amount of resources are wasted which results in low throughput and transaction failures. In this paper, we proposed ConChain, a novel blockchain scheme that combines transaction parallelism and an intelligent dependency manager to minimize conflicting transactions in blockchain networks as well as improve performance. ConChain is also capable of ensuring proper defense against major attacks due to contention.

\end{abstract}


\section{Introduction}
Blockchains, as distributed systems, require fault tolerance for record-keeping without a central authority. State Machine Replication (SMR) synchronizes servers for fault tolerance \cite{vukolic2016quest}. To handle malicious nodes, Byzantine Fault Tolerant (BFT) protocols are used for consensus \cite{buchman2016tendermint}. However, BFT struggles with high contention workloads, where conflicting transactions hinder consensus. While many researchers have worked to mitigate contention issues in classic distributed systems \cite{salehi2014contention, kostin2000randomized}, very few articles have addressed contention issues in blockchain systems. XOX Fabric \cite{gorenflo2020xox} and Fabric CRDT \cite{fabricCRDT} proposed updated frameworks that can perform better with contention workloads, but no solution to the contention problem was proposed. In this paper, we propose a novel scheme for a contention-free blockchain, ConChain. ConChain enhances ordering using a transaction dependency manager, grouping and scheduling transactions to prevent conflicts. ConChain employs parallel processing for improved performance and defends against major blockchain attacks. 

The following are the major contributions of this work:\\
1) We formally defined and simulated contention in a private blockchain.\\
2) We applied three naive solutions for reducing contention and then compared their performance.\\
3) We proposed the architecture of ConChain, a scheme that can ensure contention-free transactions, increase throughput, and defend against some well-known attacks.\\
4) We presented an outline of how ConChain will be able to defend against four major attacks.

\section{Related Works}
The popularity of cryptocurrencies and decentralized applications is leading to increasingly complex and large-scale blockchain systems. Handling the growing number of transactions securely and efficiently has prompted researchers to explore concurrency through parallel processing. Amiri et al. introduced the ParBlockchain framework, demonstrating how parallel processing can enhance transaction speed and scalability in private blockchain networks \cite{amiri2019parblockchain}. However, simultaneous processing of numerous transactions results in "conflicting transactions" or "contentions." Contention, a well-known issue in distributed systems, has been extensively studied by Salehi et al. \cite{salehi2014taxonomy}. While various solutions exist for classical distributed systems, none specifically address contention in both private and public blockchain systems. Consequently, developing a contention-free transaction framework for blockchains remains a critical task.

\section{Simulating Contention}
To understand contention, we simulated it using Hyperledger Fabric and the SmallBank dataset.

\begin{figure}[ht]
\centering
\includegraphics[height=1.0in]{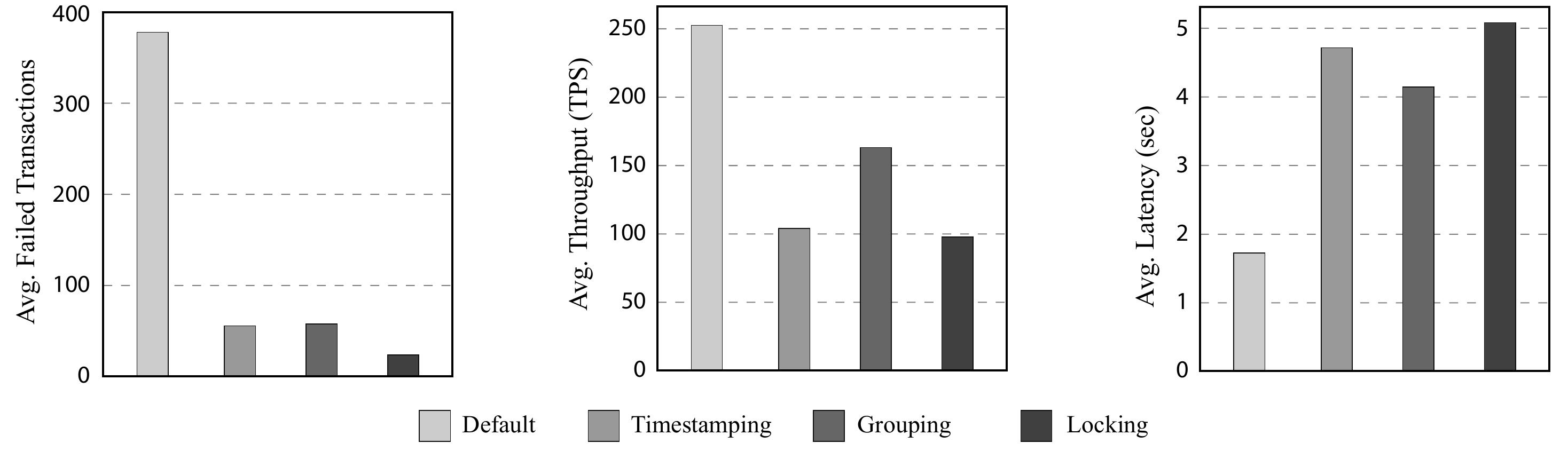}
        \caption{Comparison of different simulation results} \label{fig:sim}
\end{figure}

\textbf{Timestamping\cite{kostin2000randomized}.} Using timestamps with transactions maintained order in a "first-come, first-served" manner, as shown in Figure \ref{fig:sim}. While effective in reducing contention, this doesn't ensure contention-free transactions.

\textbf{Grouping Transactions\cite{xu9317791}.} Similar to timestamping, transactions were selected based on type (Read/Write). Prioritizing read operations reduced latency, but contention remained high in write and update operations.

\textbf{Locking\cite{xuLocking}.} Applying locks before the ordering process reduced contention significantly, as shown in Figure \ref{fig:sim}. However, this mechanism increased system latency.

\section{ConChain Architecture}
\begin{figure}[!h]
\centering
\includegraphics[width=3.5in]{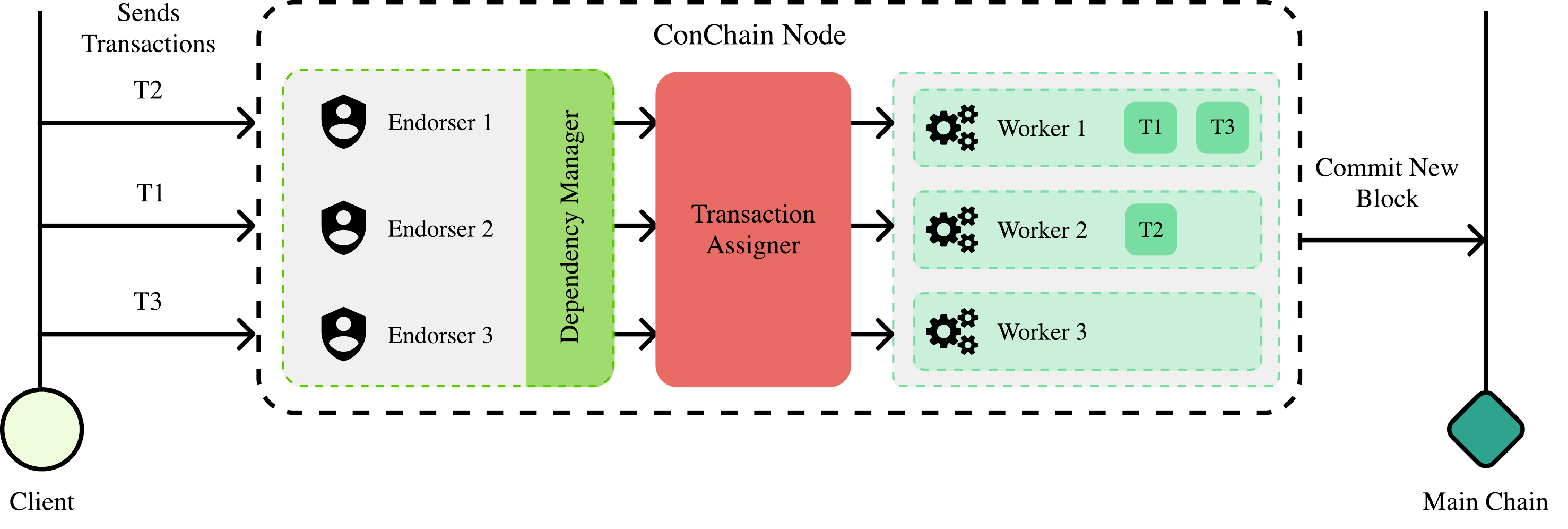}
        \caption{Architecture of ConChain} \label{fig:arch}
\end{figure}

In our simulation analysis, we discovered that modifying the ordering scheme or introducing an additional layer can effectively mitigate contention; however, this often results in a tradeoff with increased latency and reduced system throughput. ConChain, as depicted in Figure \ref{fig:arch}, addresses this challenge by incorporating two additional layers designed to facilitate contention-free transactions. The "Dependency Manager" critically assesses each transaction, assigning variables such as $readWallets$ and $writeWallets$ to indicate the relevant operations on wallets. Concurrently, the transaction assigner assesses dependencies, assigns available workers, and orchestrates conflict-free execution.

Moreover, ConChain integrates parallel processing to prioritize non-conflicting transactions, significantly enhancing the overall rate of successful transactions per second. In practical terms, this means that conflicting transactions are efficiently put on hold, allowing the system to process non-conflicting ones first. This strategic use of parallel processing is particularly impactful in scenarios where contention is prevalent, leading to a notable increase in overall system throughput.

\begin{table}[ht]
\resizebox{0.49\textwidth}{!}{%
\begin{tabular}{|c|c|c|c|c|c|c|c|}
\hline
\textbf{Scheme} &
  \textbf{Type} &
  \textbf{Nodes} &
  \textbf{Succ} &
  \textbf{Fail} &
  \textbf{Latency(s)} &
  \textbf{TPS} &
  \textbf{Succ. TPS(\%)} \\ \hline
ConChain & R           & 4 & 9974 & 26   & 0.01 & 132.5 & 88.60\% \\ \hline
Fabric   & R           & 4 & 9560 & 440  & 0.21 & 16.1  & 57.43\% \\ \hline
ConChain & RW & 4 & 8540 & 460  & 2.63 & 29.9  & 83.96\% \\ \hline
Fabric   & RW & 4 & 8126 & 1874 & 2.39 & 13.2  & 76.43\% \\ \hline
\end{tabular}%
}
\caption{Performance Comparison between ConChain and Default Hyperledger Fabric }
\label{tab:perf-comp}
\end{table}

Our simulation with 9000 transactions shows a notable reduction in failed transactions, consistently achieving over 80\% success rate compared to the default Hyperledger Fabric. The use of parallel processing also enhances overall throughput. The results, as summarized in Table \ref{tab:perf-comp}, showcase the effectiveness of ConChain.

\section{Defense Against Attacks}
ConChain effectively defends against potential attacks exploiting vulnerabilities arising from contentions.

\textbf{Double Spending Attack.} ConChain safeguards against double spending attacks by utilizing a "Dependency Manager" that rejects unrelated fake transactions, preventing delays and ordering issues caused by conflicting transactions.

\textbf{Block Withholding Attack.} To counter Block Withholding attacks, ConChain enables miners to identify fake transactions through a dependency tracker, preventing their assignment to workers and quickly recognizing and ceasing the mining of attacker-generated fake transactions.

\textbf{Balance Attack.} ConChain detects fake transactions before ordering, thwarting balance attacks attempting to fork the main chain and enabling nodes to prevent the creation of a longer, illegitimate chain.

\textbf{DDoS Attack.} ConChain defends against DDoS attacks by maintaining a transaction queue with a time limit, preventing the execution of synthetically generated conflicting transactions and thwarting the attack's success.

\section{Conclusion}
This work aims to create a contention-free, efficient, and attack-resilient blockchain scheme. Our proposed architecture, ConChain, ensures contention-free transactions by adding an extra layer to the ordering process, addressing conflicting transactions. ConChain is designed to defend against major attacks resulting from contention. We anticipate that implementing ConChain will mitigate contention in blockchain networks, enhance performance, reduce resource wastage, and provide additional defense against major attacks.

%\printbibliography
\bibliographystyle{IEEEtran}
\bibliography{IEEEabrv,references}

\end{document}